\documentstyle[twocolumn,aps,pre,psfig]{revtex}

\def\figsiz1{5cm}
\def\frontmatter{}

\begin{document}

\title{Decaying of Phase Synchronization - A Physiological Tool}

\author{%
Avi Gozolchiani$^1$,
Shay Moshel$^1$,
Jeffrey M. Hausdorff$^2$,
Ely Simon$^2$,
J\"{u}rgen Kurths$^3$,
Shlomo Havlin$^1$
}

\address{$^1$Minerva Center and Department of Physics, Bar-Ilan University, Ramat-Gan 
52900, Israel}
\address{$^2$Movement Disorders Unit, Tel-Aviv Sourasky Medical Center, Tel Aviv, Israel}
\address{$^3$Institute of Physics, Potsdam University}
\maketitle
\begin{abstract}
We describe the effects of the asymmetry of cycles
and
non-stationarity in time series on the phase synchronization method. We develop a
modified method that overcomes these effects and apply this method to study parkinsonian
tremor. Our results indicate that there is synchronization between two different hands
and provide information about the time delay separating their dynamics. 
These findings suggest that this method may be useful for detecting and quantifying 
weak synchronization between two non-stationary signals.
\end{abstract}

\pacs{PACS numbers: }

\frontmatter

\section*{\bf Introduction}
Analysis of the synchronization between subsystems of a complex dynamical system is important
for characterizing the system. Much effort has recently been focused on detecting and quantifying synchronization within non-stationary noisy
systems (see e.g., \cite{pikovsky97,kurthsRev}). In such systems, the traditional cross-correlation technique may not assure appropriate detection in all cases of interdependency ~\cite{kurthsRev}. The phase synchronization method has been found useful in identifying synchronization in several systems. It has been used to detect brain activity associated with parkinsonian tremor \cite{tass}, to describe cardiorespiratory systems ~\cite{stefa,Kotani}, and to find relations between temperature and precipitation in different regions ~\cite{diego}. 

Here we further develop the synchronization decay method and apply it
to clarify some questions related to parkinsonian tremor. Using
the accelerometric and electromyographic (EMG) records of the hands during "rest tremor", we show that the two hands' tremor movements are synchronized and that the coupling mechanism has a broad range of time response scales starting at scales that are below the detection of the measuring apparatus (the sampling rate is 400Hz) and reaching 0.5 seconds and above.                                           

\section*{\bf Effects of Asymmetry and Non-stationarity}
 
Direct application of the phase synchronization method \cite{kurthsRev}
may encounter difficulties regarding consistent asymmetry of the density of
points in each cycle, as discussed below.
The phase synchronization method uses the concept of ``analytic
signal'' (AS)~\cite{gabor46} which is useful for detecting the phase
in periodic and semi-periodic signals ~\cite{kurthsRev}. By computing
the AS $S_{A}(t)=S(t)+iS_{h}(t)$ (where $S$ is the original signal,
and $S_{h}$ is its Hilbert transform, see formulation in the
\emph{method} section) of two simultaneous records $S_{a}(t)$ and
$S_{b}(t)$, and evaluating the phase differences series
$\psi(t)=\phi_{a}(t)-\phi_{b}(t)$ (where the phase $\phi_{a,b}$ is the
angle of the AS in complex space) one can generally estimate how strongly
the signals are coupled, observing the variation of $\psi$. However, if
the analytic signals are not folded equally over the cycle , their phases
tend to stay longer in the more folded areas (see
Fig.~\ref{timering}a, for an example of asymmetric folding). In this
case, synchronized dynamics, where the 
completion of a cycle in $S_a$ depends on the completion of a cycle in
$S_b$, would be hard to detect. This difficulty arises because $S_a$ may complete a very
folded area 
(and remain in a very narrow region in phase space) while $S_b$
completes a much less folded area (and therefore swaps a wide angle in
the complex space). Thus, the phase differences series
$\psi$ ~\cite{pikovsky97} will not have a very dominant phase
lag, and the synchronization might appear to be 
insignificant. The geometry of the AS might have, therefore, a significant
impact on the ability to detect synchronization even when the dynamics of
$S_a$ and $S_b$ are coupled.  

Non-stationarity of $S_a$ and $S_b$ may lead to another difficulty. It
is accepted that the global average of the AS must be removed
before extracting the phase series, otherwise, the AS will
appear to be "folded asymmetrically" even if it is a perfect
circle (because the circle is not centered at the origin). If the
center of the AS ring is non-stationary, 
this would be the case even
after one removes the global average. A partial solution will be to
identify small regimes, where the center 
of the ring does not wonder too far, remove the average (detrending),
and calculate the synchronization index in each regime. Finally we average the synchronization
index~\cite{tass} over all regions. This compromise may still imply quite small synchronization indices when the signal is highly non-stationary. Non-stationarity will therefore also have high influence on the synchronization index.

The original work of Huygens ~\cite{huygens} discussed the
synchronization of pendulums due to interactions through the embedded media. In this case, a typical response time exists. The dynamics of one oscillator will generally follow the other oscillator in a time delayed manner. One, then, should expect that choosing the right time delay $\tau$ will improve the Synchronization index of the signals $S_a$(t) and $S_b$(t-$\tau$). 

We therefore describe a procedure (also used partially in
~\cite{diego}) with a related new index, which detects interdependencies
between processes, taking into account asymmetry and non-stationarity,
and providing a way to estimate the typical time delay ~\cite{laura}. We will assume
that interdependencies can be observed from phase differences alone,
and that the process of mutual modification does not repeat itself
exactly but has a stochastic component.

\section*{\bf The method}

In order to evaluate the significance of the synchronization indices
measured, we compare it with the "natural" values of the index
obtained when the signals are not coupled. We therefore use the
concept of ``the synchronization decay''~\cite{diego}. This concept
assumes that since the coupling is local in time, finding the
synchronization of the first time series with the future (past) of the
second time series will give a synchronization value that is related
to the geometric structure of the signal, and not to the influence of
the signals on each other. The decay of synchronization is, therefore,
just a plot of the ``simple synchronization'' indices (as defined below) as a function of
the time differences between the records. The algorithm we use to
estimate the decay of synchronization is as follows: 

Firstly we choose a reasonable time shift $\tau_{max}$ within which we
can assume the local influence of one oscillator on the other is
negligible. We set the time shift $\tau$ to be $-\tau_{max}$. If
$\tau_{max}=0$ and only one window (see below) is taken, the method is
equivalent to the standard phase synchronization method. 
\begin{enumerate} 
\item Take the signals $s_{1}(t)$, $s_{2}(t-\tau)$ when $$t\in
\left\{\begin{array}{ll}
\left[0,N-\tau_{max}-1\right] & \mbox{ if $\tau<0$ } \\
\left[\tau_{max},N-1\right] & \mbox{ if $\tau\geq0$ } 
\end{array}
\right.$$
\mbox {where} $ \tau\in[-\tau_{max},\tau_{max}]$.
\item Evaluate the analytic signals
  $$S_{A}^{\tau}(t)=S^{\tau}(t)+iS_{H}^{\tau}(t)=S^{\tau}(t)+\frac{i}{\pi
  q}*S^{\tau}(q)$$ where the superscript $\tau$ represents the time shift
  that is taken between the signals, and ``*'' is a convolution with the integration
  variable $q$. Evaluation is done using the AS form in the frequency domain\\
$$
\begin{array}{rl}
\label{analyticFour}
\displaystyle{\int} & dt\;e^{-2\pi ift}\left( S(t)+i\int dq\frac{S(q)}{\pi(t-q)}\right) =\\[10pt]
\displaystyle{\int\!\!\int} & dt\;dq\;e^{-2\pi ift}S(q)\left(\delta(t-q)+\frac{i}{\pi(t-q)}\right) =  \\[10pt]
& \left(\delta(q)+\frac{i}{\pi q}\right)*S(q) =  
 (1+sgn(f))\mbox{\bf S}(f)  = \\[15pt]
& 2\Theta(f)\mbox{\bf S}(f) 
\end{array}
$$
where $ \Theta (x)= \left\{ \begin{array}{lr}
0 & \mbox{where $x \leq 0$} \\
1 & \mbox{where $x > 0$}
\end{array}
\right.$ .
\item The analytic signals can be represented in the form $S_{A}^{\tau}(t)=A(t)\exp(i\phi^{\tau}(t))$. The phase series
$\phi^{\tau}(t)$ can be easily obtained by the trigonometric relation $$\phi^{\tau}(t)=\arctan \left(\frac{S_{H}^{\tau}(t)}{S^{\tau}(t)}\right)+\frac{\pi}{2}(1-sgn(S^{\tau})).$$ 
\item Calculate the series $$\psi_{mn}^{\tau}(t)=_{mod 2\pi}m\phi_{1}^{\tau}(t)-n\phi_{2}^{\tau}(t)$$ where $m$,$n$ are integers ($\neq0$). The ratio m:n equals to the ratio between the dominant frequencies $\omega_{2}:\omega_{1}$. 
\item Evaluate the probability density $P$ of $\psi_{mn}^{\tau}(t)$ ,
  by choosing a fine division of $[0,2\pi]$ to bins ~\cite{otnes}, and
  evaluating the probability $P_{i}$ of finding $\psi_{mn}^{\tau}(t)$
  within the i-th bin.~\cite{locks}
\item Evaluate the Shannon entropy~\cite{shannon48} of the probability histogram  $$S^{\tau}=-\sum_{i} P_{i}\log P_{i}$$ Derive the synchronization index $\rho$ ~\cite{tass} by scaling the entropy $S^{\tau}$ to be in $[0,1]$ where $0$ stands for lack of synchronization and $1$ stands for full synchronization, using the relation $$\rho^{\tau}=\frac{S_{max}-S^{\tau}}{S_{max}}$$ where $S_{max}=\log N$.
\item Set $\tau=\tau+1$, return to step 1.  
\end{enumerate}
The algorithm should be applied on relatively short time scales so
that the slips of the center of the AS' ring can be
ignored. Nevertheless the time window  should be wide enough to
contain many cycles (so that edge numerical effects can be
ignored). The decay of the synchronization index for all time windows
should be averaged over all windows. A detailed picture of the
development of coupling in time can be achieved by observing the
individual time windows.  

A "synchronized state"  is a local state. It occurs when the
synchronization index vs. time shift $\tau$ has two significantly
falling tails and one or a few maxima (e.g., see Figure~\ref{synch1111}). 
 A well pronounced synchronization occurs when the difference between the tails and the maxima is more than the standard deviation of the tails. We quantify the significance by evaluation of $$\frac{<\rho_{k\in\, middle}>-<\rho_{k\in\, tails}>}{\sqrt{<\rho_{k\in\, tails}^{2}>-<\rho_{k\in\, tails}>^{2}}}$$  

\section*{\bf Application to tremor in  Parkinson's disease}

Parkinsonian rest tremor accelerometric records, being non-stable,
gave rise to an interesting question, where the synchronization decay
diagrams might give some indication for the answer. We analyzed 29
records, from 3 male and 2 female patients, with ages between 55 and 69. Each record has a duration
of 330 seconds, and a sampling rate of 400Hz.     

The Parkinsonian rest tremor accelerometric and EMG
records in different extremities have been found to be non coherent
~\cite{bergman}, despite their concurrent motion. It can be shown that
the low coherence between two time series imposes very low
Synchronization Index as well. "Synchronization Decay" diagrams can indicate that
although the Synchronization Indices are indeed low, the hands'
dynamics might be coupled. We find that synchronization between the
hands appears to be maximal with a time delay which suggest a non
mechanical coupling (assuming that mechanical coupling time scale is
very different from neuronal coupling time scale).


We tested whether coupling between the dynamics of the extremities
exists. There is evidence in human hand tremor~\cite{sapir} and in
neuronal activity of MPTP monkey models~\cite{bergmanModes} that the first
two modes of oscillations (e.g., 4 and 8 Hz) exhibit different dynamics which can not
always be described as harmonics. We therefore computed the
synchronization decay between the hands and  
between each of the modes of the left hand and each of the modes of
the right hand. When asserting phase synchronization decay between the two full signals, one can refer to the phases extracted as ''effective phases'', and to the phases of the individual modes as ''phases''. We attribute interdependence for cases in which the decay
of synchronization is larger than 1.5 standard deviations of the tails. We found that out of
29 records of the entire signal, 19 were synchronized. The first mode of the tremor hand
was synchronized with 
the first mode of the relatively healthy hand in 16 of the records, and with its
second mode in 11 cases. The second mode of the tremor hand was
synchronized with the first mode of the relatively healthy hand in 6 cases, and
with its second mode in 7 cases. These results suggest that different
extremities oscillate dependently. A close look at Fig~\ref{signi}
reveals that synchronization of the first mode of the tremor
hand and any of the modes of the relatively healthy hand is in some cases more
sharp (above 3 standard deviations) than the synchronization of the
second mode of the tremor hand with any of the modes of the relatively healthy
hand. The appearance of a significant interdependence between
oscillators, when their coherence is insignificant~\cite{bergman}, is
a surprising result of the analysis. The other possibility, that
coherence can exist without synchronization,  has been recognized before~\cite{kurthsRev}. 
The reason for this new possibility is due to our different (less strict) demand for observing inter-dependency. We suggest that it is sufficient to demand that synchronization decays as time shift grows. 

We assume above that our findings are related to the
disease state, and that the outcoming mechanical oscillation is
indeed a reflection of the neuronal activity. However, an alternative explanation
is that the mechanical system itself (the hands and the body) causes
the relations found and not the pathology. Thus, we
must exclude the possibility that the coupling we tracked can be
fully explained by the mechanical coupling between the hands through
the body media. Our suggestion that it is due to neuronal activity is
supported by three results: 
(a.) The range of response times (between times less than 2 msec and up to
0.5 sec) that was found is recognized as typical neuronal time scale. 
(b.) In two of the cases, there was an order of magnitude difference
between the response time of second mode (in both hands) and the
response time between the first mode in one hand and the second mode
in the other. Thus, even the same person, with the same mechanical
parameters shows extremely different response times (which also
supports the claim that there are indeed several sources which rule
each hand, see~\cite{sapir}), and therefore , a mechanical coupling can not explain such
delays.
(c.) We also tested the synchronization decay between the
EMG signals. The EMG records are quite noisy, and in most
cases analysis gave no sign of the synchronization such as that found
between the accelerometric records. However there were 6 (out of 29)
records which did show good synchronization decay. Such
interdependence is better explained by the interdependence of neuronal
activity rather than mechanical couplings (because we measure the
signal from the brain, and not the actual movement).

\section*{\bf Discussion}

{\emph A.} {\bf Oscillations in the synchronization decay plot}

It is worth noting that one can usually identify the artifacts of 
non-stationarity by looking at a synchronization decay plot. For
example, in Figure~\ref{synch1111}b one can see that the
synchronization index does fall dramatically. The falling profile is
modulated with an oscillation. This oscillation can be explained with
a reasoning similar to that we gave for the index falling due to
geometry in the discussion above. When shifting the records so that
folded areas fall together most of the time, one obtains "better" synchronization. When folded area of one record overlaps the unfolded areas of the other record (and vice versa, of course), the index falls. Therefore the index falls and grows alternately as we shift the records (in the process described above). Oscillation, is therefore, due to geometry. 

{\emph B.} {\bf Why two perfect sines are not synchronized?}

 Synchronization of two sines (or of two
 pendulums with the same rod length), using our new general
 restriction (that existence of synchronization is indicated by the
 fact that the cycles of one oscillator are less coherent with the
 future cycles of the other oscillator than they are with its
 present), is very low (because there will be no change in the
 synchronization index between different time shifts). Synchronization
 means therefore a process of local (in time) feedbacks.

A simple illustration for this point is given in Fig.~\ref{tria}. A triangular wave is given by: 
$$
\begin{array}{ll} 
u(t)& ={\displaystyle \frac{2h}{\tau}} \left\{ \sum\limits_{i=0}^{\infty}[sgn(-t+(1+2i){\displaystyle\frac{\tau}{2}})-\right.\\ \\[5pt]
&\left. sgn(-t+i\tau)] (t-(0.5+2i){\displaystyle\frac{\tau}{2}})+ \right. \\ \\[5pt] 
& \left. [sgn(-t+(2+2i){\displaystyle\frac{\tau}{2}})-\right.\\ \\
& \left. sgn(-t+(1+2i){\displaystyle\frac{\tau}{2}})] (-t+(1.5+2i){\displaystyle\frac{\tau}{2}}) \right\}, 
\end{array}
$$
where ${\displaystyle\frac{h}{2}}$ is the amplitude, and $\tau$ is the period. A decay of synchronization index between the two first modes of oscillation of a perfect triangular periodic wave does not exist because each cycle of the first mode is prefectly synchronized with each cycle of the second mode, and not just with the relatively close cycles ($\tau$ is not a function of time, therefore each cycle is identical in its phase dynamics with previous cycles). Each point on the decay of synchronization plot is a local maximum.

Even if you perturb by adding ($u'(t)=u(t)+\epsilon \eta (t)$ when
 $\eta (t)$ is some white noise) or multiplying ($h=h(t)$) the wave by
 some random noise, you won't get decay of the synchronization index,
 because phases stay the same (detection of the phases might differ a
 bit, but not consistently enough to reduce significantly the
 Synchronization Index). In Fig.~\ref{tria}a,b,c there is a
 superposition of several noisy triangular waves ($\tau$ is fixed for
 each of them, and $h=h(t)$). It's obvious that the Synchronization
 Index is not 1 all over the diagram (because detecting is not
 perfect) but there's no decaying (Fig.~\ref{tria}c). Nevertheless, perturbing the
 frequency~\cite{gentle} ($\tau=\tau (t)$\. 
  Fig.~\ref{tria}d,e,f)
 mimics a dependency between the two modes of oscillation, because a
 local change in the proceeding of one mode is seen simultaneously in
 the other mode. In real systems, such "simultaneous" local changes
 are the heart of a dynamical synchronization process. This simple
 example suggests that harmonies of a signal with changing local
 period synchronize (in the sense of decaying synchronization) in
 order to build the wave form. Indeed, the decaying of the
 synchronization index plot in Fig.~\ref{tria}f shows sharp
 decay in both tails. Simultaneous local (in
 time) changes in two time series (and therefore - the measure of
 synchronization decay) indicate interdependence.    

\bigskip
 This work was supported in part by NIH grants
 NS-35069, RR-13622, HD-39838 and AG-08812 
 and by the United States-Israel Bi-National Science
 Foundation.

\def\figureI{
\begin{figure}[thb]
\centerline{\psfig{figure=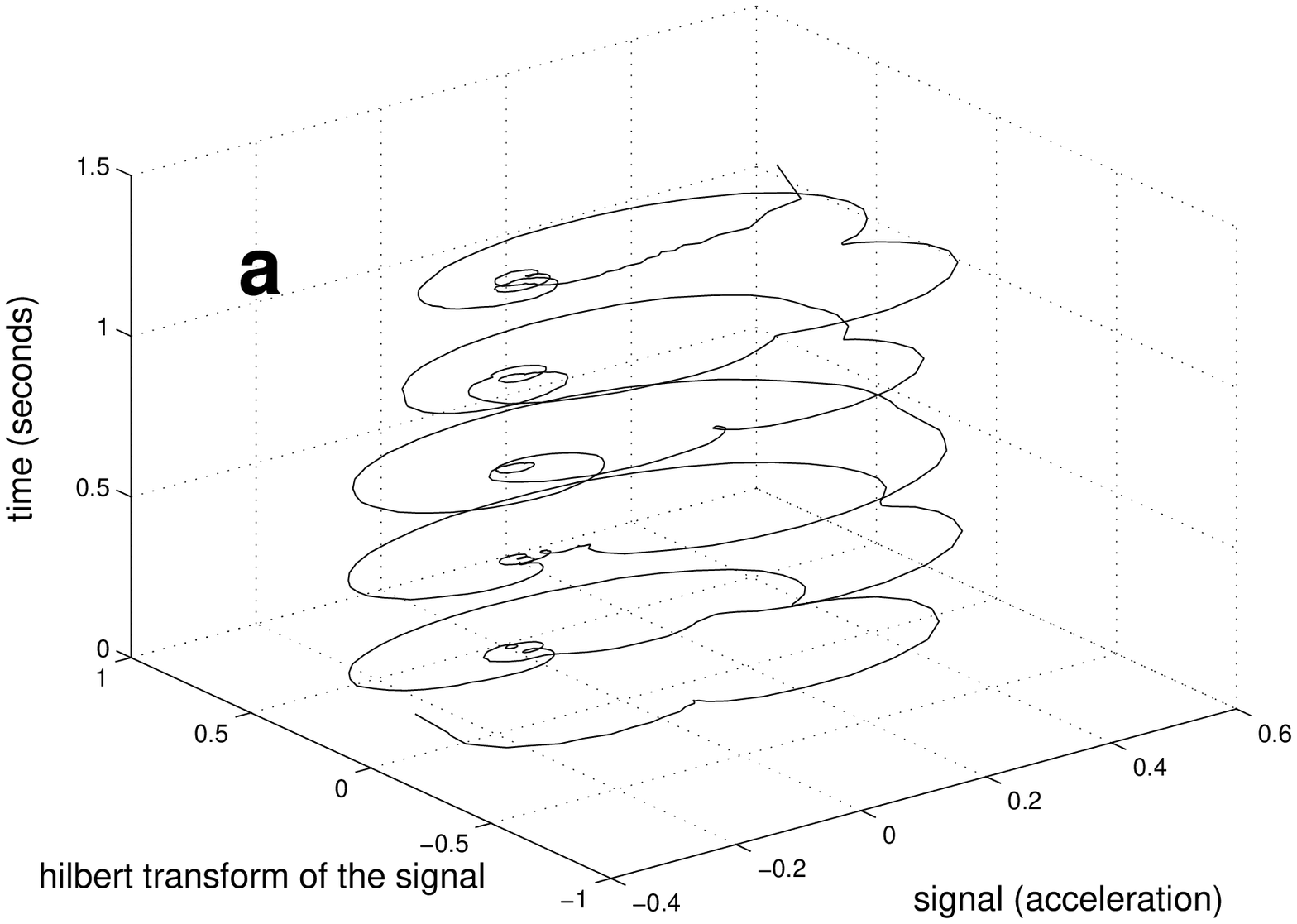,width=9cm}}
\centerline{\psfig{figure=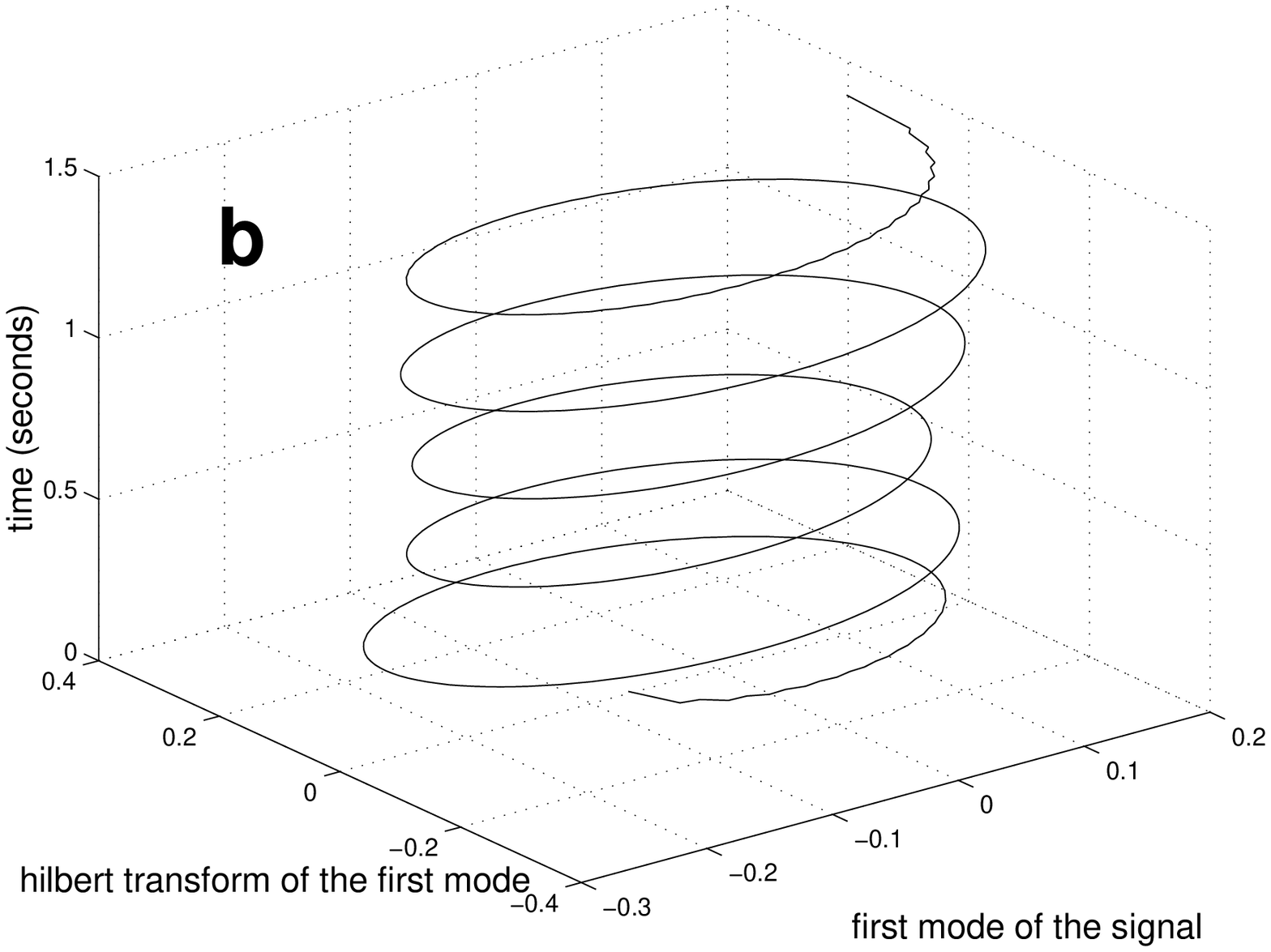,width=9cm}}
{
\vspace*{0.0truecm}
\caption{\label{timering}
{\small Example of a parkinsonian tremor record's analytic signal as a
   function of time. The analytic signal always oscillates in a
   counter clock wise direction.\\ a: the full signal's transform.\\ b:
   analytic signal of only first mode of oscillation.} 
}}
\end{figure}
}

\def\figureIII{
\begin{figure}[thb]
\centerline{\psfig{figure=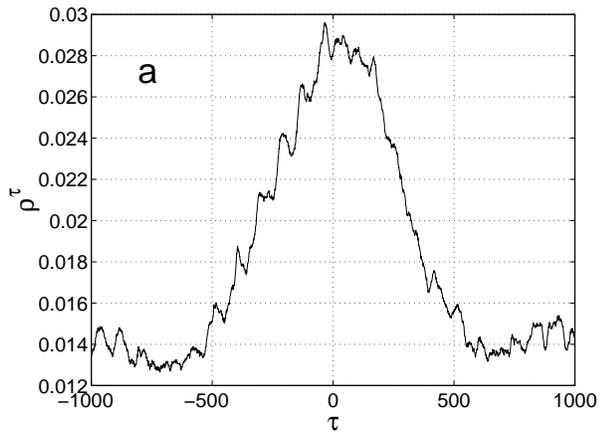,width=8cm}}
\centerline{\psfig{figure=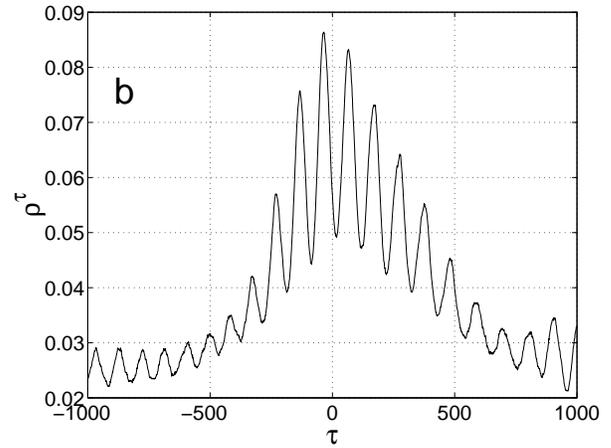,width=8cm}}
{
\vspace*{0.0truecm}
\caption{\label{synch1111} 
{\small
(a) Synchronization decay between oscillations in both hands. Records
  here are not filtered.
(b) Synchronization decay of first mode in both hands. There are
    clearly decaying tails, but the figure is modulated by a strong
    oscillation. In both cases one sees that $\rho^{0}$, the
    standard synchronization index, is extremely small, but is
    significantly larger than the background noise. 
}}
}
\end{figure}
}
\def\figureIV{
\begin{figure}[thb]
\centerline{\psfig{figure=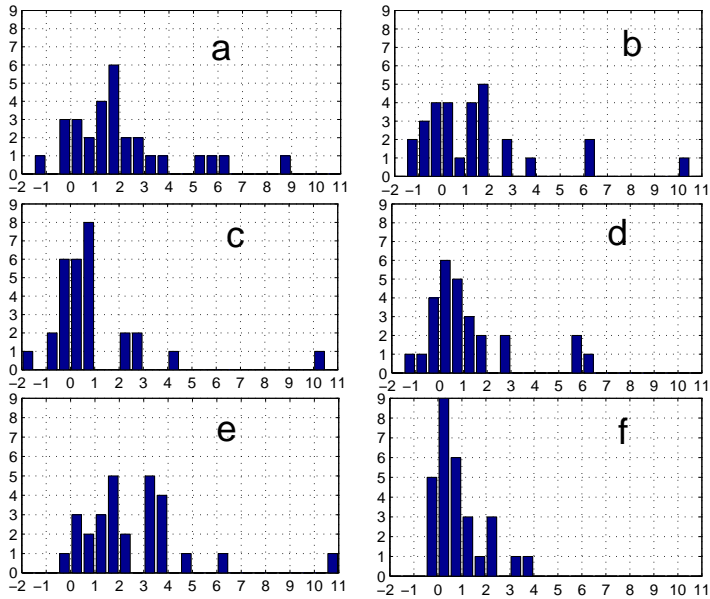,width=9.5cm}}
{
\vspace*{0.0truecm}
\caption{\label{signi} 
{\small
Distribution of difference between center and tails mean values (in
standard deviation units) in the synchronization decay diagram testing interdependency between a. First mode of healthy hand vs. first mode of tremor hand. b. Second mode of healthy hand vs. first mode of tremor hand. c. First mode of healthy hand vs. second mode of tremor hand. d. Second mode of healthy hand vs. second mode of tremor hand. e. Full Tremor signals of both hands. f. Full EMG signals of both hands.  
}
}}
\end{figure}
}
\def\figureV{
\begin{figure}[thb]
\centerline{\psfig{figure=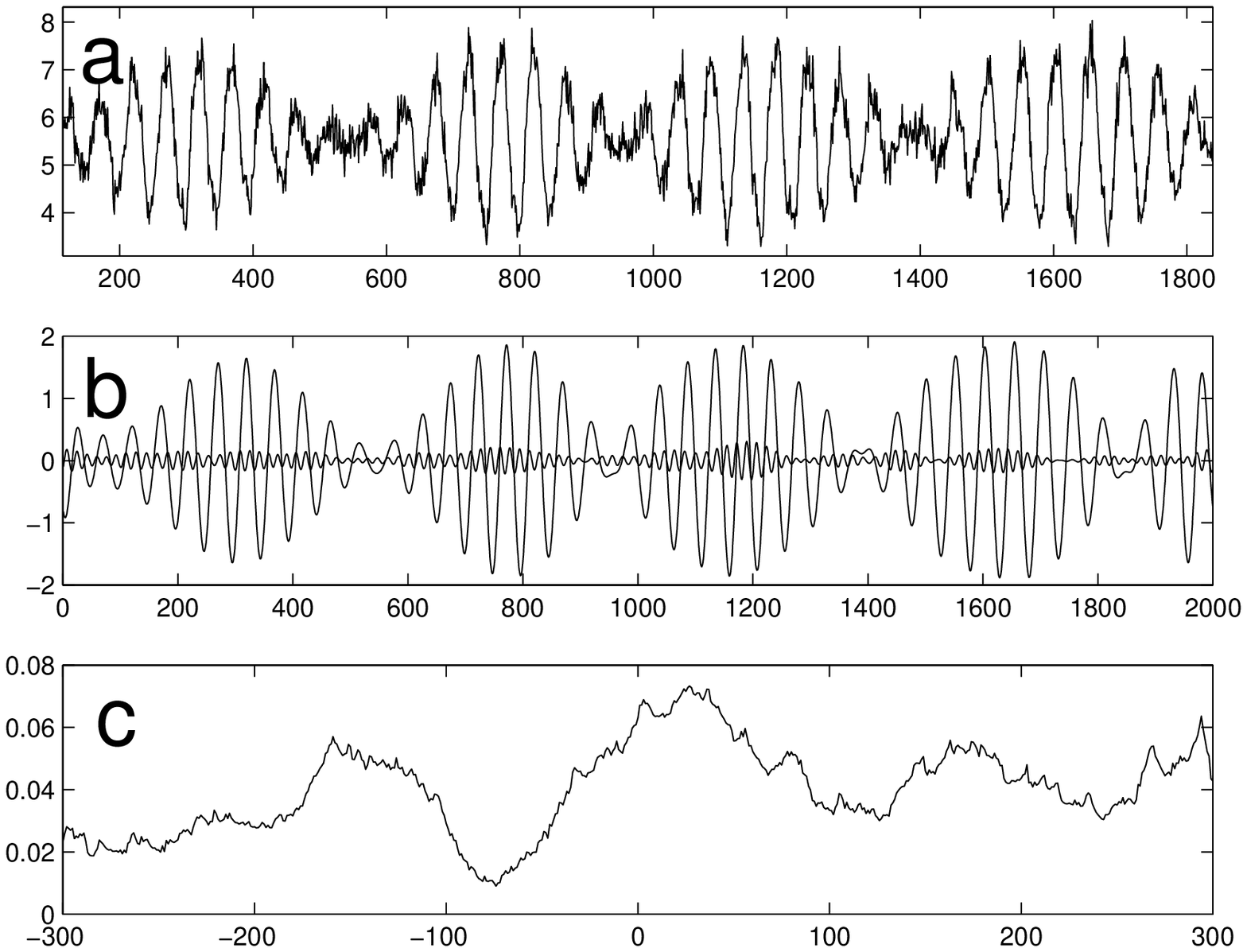,width=9cm}}
\centerline{\psfig{figure=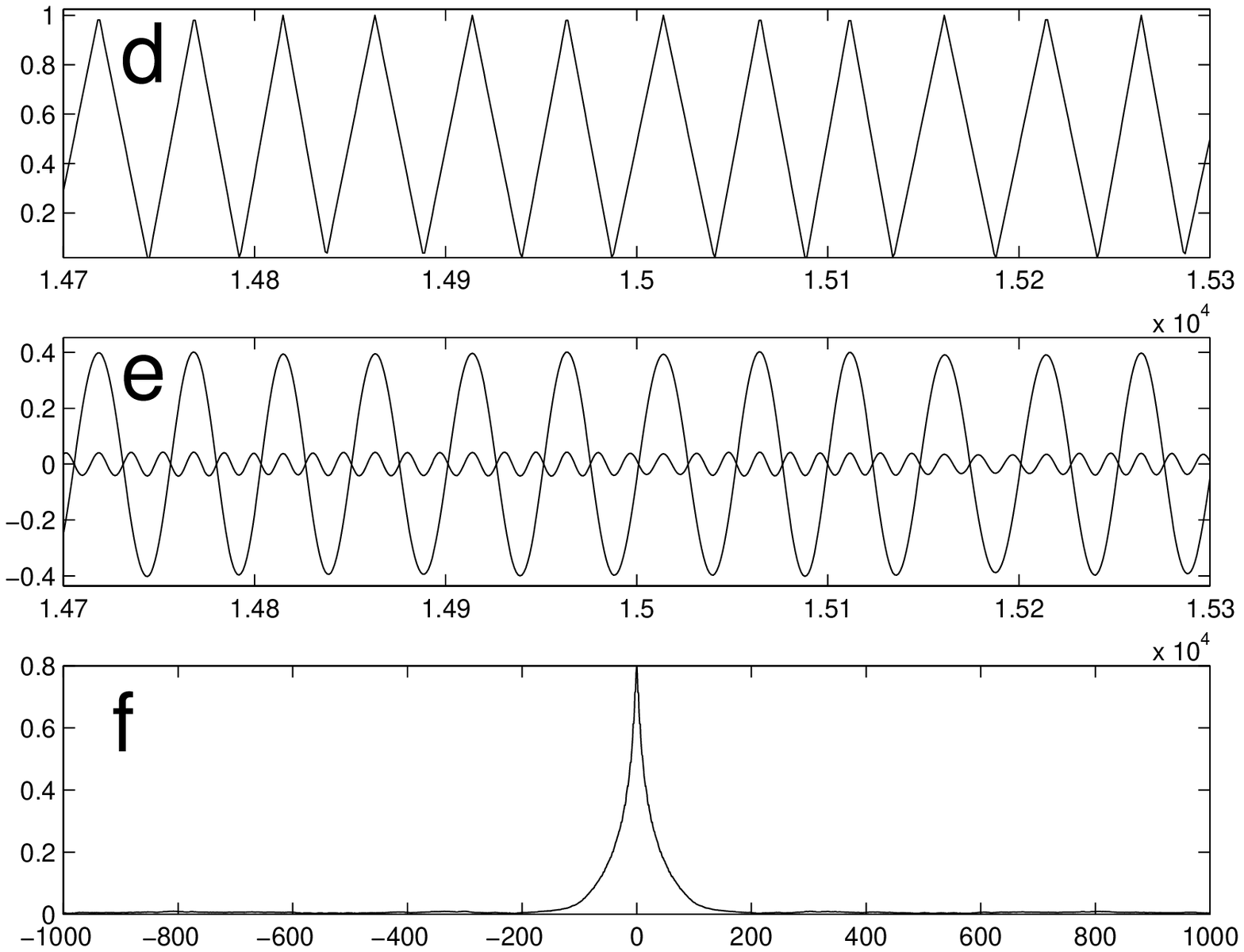,width=9cm}}
{
\vspace*{0.0truecm}
\caption{\label{tria} 
{\small
a,b,c : A superposition (with random in time coefficients) of 10 triangular waves each with slightly different wave length, the separation into two dominant modes and the Decay of the Synchronization Index plot.\\
d,e,f : A triangular wave, its separation to modes and the Decay of the Synchronization index plot. On each cycle a new wavelength is chosen randomly.
}
}}
\end{figure}
}

\figureI

\figureIII

\newpage

\figureIV

\newpage

\figureV


\begin{thebibliography}{00}
\bibitem{locks} When the two oscillators are phase-locked,
  $\psi_{mn}$ satisfies the relation\\
  $|\psi_{mn}^{0}(t)-\epsilon|<\delta$  for a certain $\epsilon$ where
  $\delta<2\pi$ for all t, thus $P_{i}=0$ for all bins outside the $\epsilon$-zone.
\bibitem{gentle} Here we modified $\tau$ once in each cycle, giving it 10 different
 discrete values alternately. The reason for this perturbing strategy
 is that the phase is a zero Lyapupnov parameter, and therefore its
 dynamic is too sensitive to frequent noise.
\end{thebibliography}
\end{document}